\documentclass[a4paper,12pt]{article}

\usepackage{cite,amsmath,amsfonts,amsthm,fullpage}
\usepackage{amssymb}
\usepackage{amsbsy}
\usepackage{epsfig}
\usepackage{verbatim}
\makeatletter \@addtoreset{figure}{section}
\def\thefigure{\thesection.\@arabic\c@figure}
\def\fps@figure{h, t}
\@addtoreset{table}{bsection}
\def\thetable{\thesection.\@arabic\c@table}
\def\fps@table{h, t}
\@addtoreset{equation}{section}

\newtheorem{corollary}{Corollary}[section]
\newtheorem{definition}{Definition}[section]

\newtheorem{proposition}{Proposition}[section]

\newtheorem{examps}{Examples}[section]

\newtheorem{lemma}{Lemma}[section]
\newtheorem{remark}{Remark}[section]
\newtheorem{remarks}[remark]{Remarks}
\def\mod{\,\hbox{mod}\,}

\def\bx{\begin{example}}
\def\ex{\end{example}}
\def\bxs{\begin{examps}. \rm\begin{enumerate}}
\def\exs{\end{enumerate}\end{examps}}
\def\bd{\begin{definition}}
\def\ed{\end{definition}}
\def\bp{\begin{proposition}\rm}
\def\ep{\end{proposition}}
\def\bc{\begin{corollary}}
\def\ec{\end{corollary}}
\def\bl{\begin{lemma}\em}
\def\el{\end{lemma}}
\def\be{\begin{equation}}
\def\ee{\end{equation}}
\def\br{\begin{remark}\rm\small}
\def\er{\end{remark}}
\def\brs{\begin{remarks}.\\ \rm\
\begin{enumerate}}
\def\ers{\end{enumerate}\end{remarks}}
\def\bea{\begin{eqnarray}}
\def\eea{\end{eqnarray}}

\def\det{\mathrm {det}}

\def\&{&{\hskip -20pt}}

\def\AA{{\mathcal A}}

\def\Zb{{\mathbf Z}}

\date{}

\begin{document}
\baselineskip 16pt
\begin{flushright}
CRM-xxxx (2005)\\
nlin.SI/05xxxxxx
\end{flushright}
\medskip
\begin{center}
\begin{Large}\fontfamily{cmss}
\fontsize{17pt}{27pt} \selectfont \textbf{Fermionic construction
of partition function for multi-matrix models and multi-component
TL hierarchy}\footnote{Work of (J.H.) supported in part by the
Natural Sciences and Engineering Research Council of Canada
(NSERC) and the Fonds FCAR du Qu\'ebec; that of (A.O.) by the
Russian Academy of Science program  ``Fundamental Methods in
Nonlinear Dynamics" and  RFBR grant No 05-01-00498.}
\end{Large}\\
\bigskip
\begin{large}  {J. Harnad}$^{\dagger \ddagger}$\footnote{harnad@crm.umontreal.ca}
 and {A. Yu. Orlov}$^{\star}$\footnote{ orlovs@wave.sio.rssi.ru}
\end{large}
\\
\bigskip
\begin{small}
$^{\dagger}$ {\em Centre de recherches math\'ematiques,
Universit\'e de Montr\'eal\\ C.~P.~6128, succ. centre ville, Montr\'eal,
Qu\'ebec, Canada H3C 3J7} \\
\smallskip
$^{\ddagger}$ {\em Department of Mathematics and
Statistics, Concordia University\\ 7141 Sherbrooke W., Montr\'eal, Qu\'ebec,
Canada H4B 1R6} \\
\smallskip
$^{\star}$ {\em Nonlinear Wave Processes Laboratory, \\
Oceanology Institute, 36 Nakhimovskii Prospect\\
Moscow 117851, Russia } \\
\end{small}
\end{center}
\bigskip
\bigskip

\begin{center}{\bf Abstract}
\end{center}
\smallskip

\begin{small}
We use $p$-component fermions $(p=2,3,\dots)$ to present
$(2p-2)N$-fold integrals as a fermionic expectation value. This
yields fermionic representation for various $(2p-2)$-matrix
models. Links with the $p$-component KP hierarchy and also with
the $p$-component TL hierarchy are discussed. We show that the set
of all (but two) flows of $p$-component TL  changes standard
matrix models to new ones.
\end{small}
\bigskip

\section{Introduction}

\

Let $d\mu_\alpha(x,y)$ be a set of measures (in general, complex),
supported on a finite set of products of curves in the complex $x$
and $y$ planes.

Let $\rho_{\alpha}$, $\alpha=2,\dots,p-1$, be a set of functions
in two variables.

Let $x^{(\alpha)}=(x^{(\alpha)}_1,\dots,x^{(\alpha)}_N)$ and
$y^{(\alpha)}=(y^{(\alpha)}_1,\dots,y^{(\alpha)}_N)$, $
\alpha=1,\dots,p$ are two sets of variables, where $x^{(p)}$ and
$y^{(1)}$ are fixed by
 \be\label{fixing-x-p-y-1}
x^{(p)}_i=y^{(1)}_i=N-i,\quad i=1,\dots,N
 \ee

We shall use the following notation
 \be\label{dmu-alpha}
d\mu_{\alpha}(x^{(\alpha)},y^{(\alpha+1)}):=\prod_{i=1}^N
d\mu_{\alpha}(x^{(\alpha)}_i,y^{(\alpha+1)}_i)
 \ee

 We consider the following
integral over $(2p-2)N$ variables
$x^{(\alpha)}=(x^{(\alpha)}_1,\dots,x^{(\alpha)}_N)$ and
$y^{(\alpha+1)}=(y^{(\alpha+1)}_1,\dots,y^{(\alpha+1)}_N)$, $
\alpha=1,\dots,p-1$:
 \be
  {Z}_N=      \label{Z_N}
\int
\prod_{\alpha=1}^{p}\varrho_\alpha(y^{(\alpha)},x^{(\alpha)})\prod_{\alpha=1}^{p-1}
d\mu_{\alpha}(x^{(\alpha)},y^{(\alpha+1)})
 \ee
 where
  \[
  \varrho_1 (y^{(1)},x^{(1)})=\det \left( \left(x_i^{(1)}\right)^{y_j^{(1)}}\right)_{i,j=1,\dots,N}= \prod_{i > j}^N
  (x^{(1)}_i-x^{(1)}_j)=:\Delta_N(x^{(1)}),
 \]
  \[
  \varrho_p (y^{(p)},x^{(p)})=\det \left( \left(y_i^{(p)}\right)^{x_j^{(p)}}\right)_{i,j=1,\dots,N}= \prod_{i > j}^N
  (y^{(p)}_i-y^{(p)}_j)=:\Delta_N(y^{(p)})
 \]
 are Vandermonde determinants, and where
 \be
 \varrho_\alpha (y^{(\alpha)},x^{(\alpha)})=\det
\left(\rho_\alpha
(y^{(\alpha)}_i,x^{(\alpha)}_j)\right)_{i,j=1,\dots,N},\quad
\alpha=2,\dots,p-1
 \ee

 Developing each
$\det\rho_\alpha $ into $N!$ monomial terms (each is labeled by an
element of the permutation group $S_N$), and, for given choice of
the element of the permutation group, say $\sigma$, using the
change of variables inside of each $N$-fold integral to the left
(namely, $x^{(\beta)}\to \sigma (x^{(\beta)})$, $y^{(\beta)}\to
\sigma (y^{(\beta)})$ for all $\beta\le\alpha$), then, using the
anti-symmetry of $\Delta_N(x^{(1)})$ (which is the integrand of
the very left $N$-fold integral), one finds that each
 term of the mentioned development yields the same contribution.
This is a standard way to
 re-write (\ref{Z_N}) as
\[
 Z_N=\frac{1}{(N!)^{2p-2}}\int \dots \int  \Delta_N(x^{(1)})
 \prod_{i=1}^N d\mu_1(x^{(1)}_i,y^{(2)}_i)\rho_2(y^{(2)}_i,x^{(2)}_i)
 \]
 \[
 \int \dots \int
 \prod_{i=1}^N \rho_3(y^{(3)}_i,x^{(3)}_i) d\mu_2(x^{(2)}_i,y^{(3)}_i)
\]
\[
\cdots
\]
\[
 \int \dots \int
 \prod_{i=1}^N \rho_{p-1}(y^{(p-1)}_i,x^{(p-1)}_i)
 d\mu_{p-2}(x^{(p-2)}_i,y^{(p-1)}_i)
\]
\[
 \int \dots \int
\prod_{i=1}^N d\mu_{p-1}(x^{(p-1)}_i,y^{(p)}_i)\Delta_N(y^{(p)})
 \]

Integrals (\ref{Z_N}) may be related to the so-called
determinantal ensembles \cite{soshnikov}.

For special choice of measures $d\mu_\alpha$ and functions
$\rho_\alpha$, integrals (\ref{Z_N}) arose in the study of
multi-matrix  models, where matrices $M_1,M_2,M_3,\dots,M_{2p-2}$
with eigenvalues respectively equal to the sets $\{x^{(1)}_i,\;
i=1,\dots,N\}$,$\{y^{(2)}_i,\; i=1,\dots,N\}$,$\{x^{(2)}_i,\;
i=1,\dots,N\}$,\dots,$\{y^{(p)}_i,\; i=1,\dots,N\}$, are coupled
in an open chain. It occurs in case when one can reduce the
integration over matrix entries to the integrals over eigenvalues
of each matrix (for these topic see \cite{ZJZ},\cite{1'} and
Appendices to \cite{OS},\cite{paper1}). Depending on $d\mu_\alpha$
and functions $\rho_\alpha$, these are models of normal matrices,
and certain
 models
 of random Hermitian (anti-Hermitian) matrices and certain models of random
 unitary matrices, see
 \cite{Mehta},\cite{IZ},\cite{ZKMMO},\cite{EM},\cite{BEH1},\cite{ZJZ},\cite{1'},
 together with discrete versions of these matrix models \cite{OS}.

For instance, to obtain the partition function for the model of
random $N$ by $N$ Hermitian matrices, $M_1,\dots,M_{2p-2}$,
coupled in a chain,
\[
\int e^{{\mbox Tr} \sum_{k=1}^{2p-2}V_k(M_k)+ {\mbox Tr} (
c_1M_1M_2+\cdots+c_{2p-3}M_{2p-3}M_{2p-2})}\prod_{k=1}^{2p-2}dM_k,
\]
 one takes
 \be\label{for-H-chain-mu}
d\mu_\alpha(x,y)=e^{c_{2\alpha-1}
xy+V_{2\alpha-1}(x)+V_{2\alpha}(y)}, \quad \alpha=1,\dots,p-1
 \ee
 \be\label{for-H-chain-rho}
\rho_\alpha(x,y)=e^{c_{2\alpha} xy}, \quad \alpha=2,\dots,p-1
 \ee
Then $x^{(\alpha)}_i$, $i=1,\dots,N$, are eigenvalues of Hermitian
matrices with odd numbers, say $M_{2\alpha-1}$, while
$y^{(\alpha)}_i$, $i=1,\dots,N$, are eigenvalues of $M_{2\alpha}$,
$\alpha=1,\dots,p-1$. For future purpose, let us use the obvious
freedom to re-write $d\mu_\alpha$ and $\rho_\alpha$ in form
 \be\label{for-H-chain-mu}
d\mu_1(x,y)\to e^{c_{1} xy+V_{1}(x)},\quad d\mu_{p-1}(x,y)\to
e^{c_{2p-3} xy+V_{2p-2}(x)},
 \ee
 \be\label{for-H-chain-rho}
d\mu_{\alpha}(x,y)\to e^{c_{2\alpha-1} xy},\;\alpha=
2,\dots,p-2,\qquad \rho_\alpha(x,y)\to e^{c_{2\alpha}
xy+V_{}(x)+V_{}(y)}, \quad \alpha=2,\dots,p-1
 \ee

\

In the present paper we have two tasks.

First, we equate the integral (\ref{Z_N}) to the fermionic vacuum
expectation value. Here we use the so-called $p$-component
fermions. This may be considered as a continuation of of the work
\cite{ZKMMO}.

Second, as a continuation of \cite{GMO}, we relate $Z_N$ to the
coupled $p$-component KP hierarchies, or, the same to the $p$
component TL hierarchy. For this purpose we consider the following
deformation of the first and the last measures
\begin{equation}\label{measure-1-defor}
d \mu_1(x,y)\to d\mu_1(x,y|{\bf t}^{(1)},n,\bar{\bf
t}^{(1)}):=x^{n_1} e^{V(x,{\bf t}^{(\alpha)})+V(x^{-1},\bar{\bf
t}^{(1)}) }d\mu_1(x,y),
\end{equation}
\begin{equation}\label{measure-p-1-defor}
d \mu_{p-1}(x,y)\to d\mu_{p-1}(x,y|{\bf t}^{(p)},n,\bar{\bf
t}^{(p)}):=y^{n_{p}} e^{V(y,{\bf t}^{(p)})+V(y^{-1},\bar{\bf
t}^{(p)}) }d\mu_{p-1}(x,y)
\end{equation}
\begin{equation}\label{V-x-t}
 V(x,{\bf t}^{(\alpha)})=\sum_{m=1}^\infty x^mt_m^{(\alpha)},\quad
  V(x^{-1},\bar{\bf t}^{(\alpha)})=
 \sum_{m=1}^\infty x^{-m}\bar{t}_m^{(\alpha)},\quad
 \alpha=1,p,
\end{equation}
and also the following deformations of functions  $\rho_\alpha$,
$\alpha=2,\dots,p-1$,
\begin{equation}\label{rho-defor}
\rho_\alpha(x,y)\to \tau_{n^{(\alpha)}}({\bf
t}^{(\alpha)}+[x],\bar{\bf t}^{(\alpha)}+[y]),\quad
\alpha=2,\dots,p-1,
\end{equation}
where in the right hand side we have  tau functions (labeled by
$\alpha=2,\dots,p-1$) of the one-component TL hierarchy, and where
$+[x]$ and $+[y]$ denote the so-called Miwa shift of a
 TL  (a one-component TL) higher times, details are written down below.

The deformation (\ref{measure-1-defor})-(\ref{rho-defor}) relates
integrals (\ref{Z_N}) to the coupled $p$-component KP hierarchies.
If in (\ref{Z_N}) we take the deformed measures and the deformed
functions $\rho_\alpha$ as described above, then, $Z_N$ turns out
to be a certain tau function of coupled $p$-component KP, or the
same, $p$-component TL hierarchy,
 where the sets of complex numbers ${\bf
t}^{(\alpha)}=(t_1^{(\alpha)},t_2^{(\alpha)},\dots)$, $\bar{\bf
t}^{(\alpha)}=(\bar{t}_1^{(\alpha)},\bar{t}_2^{(\alpha)},\dots)$,
 and the set of integers $n^{(\alpha)}$, $
\alpha=1,\dots,p$, play the role of higher $p$-component TL times.
For the sake of brevity we shall also use the notations ${\bf
t}=({\bf t}^{(1)},\dots,{\bf t}^{(p)})$ and $\bar{\bf t}=(\bar{\bf
t}^{(1)},\dots,\bar{\bf t}^{(p)})$.

Important to mark, that the deformation (\ref{measure-1-defor}),
(\ref{measure-p-1-defor}) and (\ref{rho-defor}) seems do not keep
the form (\ref{for-H-chain-mu})-(\ref{for-H-chain-rho}). In our
case the interaction
$e^{c_{2\alpha}M_{2\alpha}M_{2\alpha+1}+V_{2\alpha}(M_{2\alpha})+V_{2\alpha+1}(M_{2\alpha+1})}$
is replaced by arbitrary chosen one-component TL tau function
(\ref{rho-defor}) where $x=x^{(\alpha)}$  is the collection of
eigenvalues of the matrix $M_{2\alpha}$  while $y=y^{(\alpha)}$ is
the collection of eigenvalues of the matrix $M_{2\alpha+1}$.


Let us note that one can consider $(p-1)N$-fold integrals if he
specifies the measures $d\mu_\alpha(x,y)$ to be proportional to
Dirac delta function which equate $x$ to a function of $y$ (it may
be $\delta(x-y)$).

The present paper is a part of series of papers devoted to
fermionic approaches to multi-fold integrals, see \cite{HO1},
\cite{paper1}, \cite{paper3}. Let us mark that our fermionic
constructions of papers \cite{paper1}, \cite{paper3} and of the
present paper are different from what was considered in
\cite{ZKMMO} and also different of  \cite{HO1}.

\subsection{Free fermions \label{freefermi}}

 Let $\AA$ be the complex Clifford algebra over $ \mathbb{C}$ generated
by \emph{charged free fermions} \hbox{$\{f_i$, ${\bar f}_i\}_{i\in
{\bf Z}}$},  satisfying the anticommutation relations
\begin{equation}\label{fermions}
[f_i,f_j]_+=[{\bar f}_i,{\bar f}_j]_+=0,\quad [f_i,{\bar
f}_j]_+=\delta_{ij}.
\end{equation}
Any element of the linear part \be W:=\left(\oplus_{m \in
\Zb}\mathbb{ C}f_m\right)\oplus \left(\oplus_{m\in \Zb}\mathbb{
C}{\bar f}_m\right)
 \ee will be referred to as a {\em free fermion}. We also introduce the
 fermionic free fields
\be \label{fermions-fourier}
    f(x):=\sum_{k\in\Zb}f_kx^k,\quad
    {\bar f}(y):=\sum_{k\in\Zb}{\bar f}_ky^{-k-1},
\ee which may be viewed as generating functions for the $f_j,
\bar{f}_j$'s.

This Clifford algebra has a standard Fock space representation
defined as follows. Define the complementary, totally null (with
respect to the underlying quadratic form) and mutually dual
subspaces
 \be W_{an}:=\left(\oplus_{m<0}\mathbb{
C}f_m\right)\oplus \left(\oplus_{m\ge 0}\mathbb{ C}{\bar
f}_m\right), \qquad  W_{cr}:=\left(\oplus_{m\ge
0}\mathbb{C}f_m\right)\oplus \left(\oplus_{m< 0}\mathbb{ C}{\bar
f}_m\right), \ee and consider the left  and right $\AA$-modules
\be
 F:=\AA/\AA W_{an}, \qquad {\bar F}:=W_{cr}\AA{\backslash}\AA.
\ee These are cyclic $\AA$-modules generated by the vectors
 \be
|0\rangle= 1\  \mod \  \AA  W_{an}, \qquad  \langle 0|= 1 \  \mod
\ W_{cr}\AA , \ee respectively, with the properties \bea
\label{vak} f_m |0\rangle=0 \qquad (m<0),\qquad {\bar
f}_m|0\rangle =0 \qquad (m \ge 0) , \cr
 \langle 0|f_m=0 \qquad (m\ge 0),\qquad \langle
0|{\bar f}_m=0 \qquad (m<0) . \eea The {\em Fock spaces} $F$ and
${\bar F}$ are mutually dual, with the hermitian pairing defined
via the linear form $\langle 0| |0 \rangle$ on $\AA$ called the
{\em vacuum expectation value}.  This is determined by \bea
\label{psipsi*vac} \langle 0|1|0 \rangle&\&=1;\quad \langle
0|f_m{\bar f}_m |0\rangle=1,\quad m<0; \quad  \langle 0|{\bar
f}_mf_m
|0\rangle=1,\quad m\ge 0 ,\\
\label{end}
 \langle 0| f_n
 |0\rangle&\&=\langle 0|{\bar f}_n
 |0\rangle=\langle 0|f_mf_n |0\rangle=\langle 0|{\bar f}_m{\bar f}_n
 |0\rangle=0;
 \quad \langle 0|f_m{\bar f}_n|0\rangle=0, \quad m\ne n,\cr
 &\&
\eea together with the Wick theorem which implies, for any finite
set of elements $\{w_k \in W\}$, \bea \label{Wick} \langle 0|w_1
\cdots w_{2n+1}|0 \rangle &\&=0,\cr
 \langle 0|w_1
\cdots w_{2n} |0\rangle &\&=\sum_{\sigma \in S_{2n}} sgn\sigma
\langle 0|w_{\sigma(1)}w_{\sigma(2)}|0\rangle \cdots \langle 0|
w_{\sigma(2n-1)}w_{\sigma(2n)} |0\rangle . \eea Here  $\sigma$
runs over permutations for which $\sigma(1)<\sigma(2),\dots ,
\sigma(2n-1)<\sigma(2n)$ and $\sigma(1)<\sigma(3)<\cdots
<\sigma(2n-1)$.

Now let  $\{w_i\}_{ i=1,\dots,N}$, be linear combinations of the
$f_j$'s only, $j\in\Zb$, and  $\{{\bar w}_i\}_{ i=1,\dots,N}$
linear combinations of the ${\bar f}_j$'s, $j \in\Zb$.
Then(\ref{Wick}) implies
\begin{equation}\label{Wick-det}
\langle 0|w_1\cdots w_{N}{\bar w}_N \cdots {\bar w}_1 |0\rangle
=\det\; (\langle 0| w_i{\bar w}_j|0\rangle)\ |_{i,j=1,\dots,N}
\end{equation}

Following refs. \cite{DJKM1},\cite{JM},  for all $ N\in \Zb$, we
also  introduce the states
 \be \label{1-vacuum}
  \langle  N|:=\langle 0|C_{N}
 \ee
where
 \bea
\label{1-vacuum'} C_{N}&\&:={\bar f}_0\cdots {\bar
f}^{(\alpha)}_{N-1}
 \quad {\rm if }\ N>0 \\
C_{N}&\&:={ f}_{-1}\cdots { f}_{N}
\quad {\rm if}\ N<0  \\
 C_{N}&\&:=1 \quad {\rm if}\ N=0
 \eea
and
 \be \label{1-vacuum-r}
    |N \rangle:={\bar C}_{N}|0\rangle
 \ee
where
 \bea
\label{1-vacuum'-r} {\bar C}_{N}&\&:=f_{N-1}\cdots
f_0 \quad {\rm if }\ N>0 \\
{\bar C}_{N}&\&:={\bar f}_{N}\cdots {\bar f}_{-1}
\quad {\rm if}\ N<0  \\
 {\bar C}_{N}&\&:=1 \quad {\rm if}\ N=0
 \eea
The states   (\ref{1-vacuum}) and (\ref{1-vacuum-r}) are referred
to as the left and right charged vacuum vectors, respectively,
with charge $N$.

In what follows we use the notational convention
\begin{equation}\label{Delta(N)}
\Delta_N(x)=\det \; (x_i^{N-k})|_{i,k=1,\dots,N}\ (N>0),\quad
\Delta_0(x)=1,\quad \Delta_N(x)=0\ (N<0).
\end{equation}
From the relations
 \be
 \langle 0|  {\bar f}_{N-k} f(x_{i})|0\rangle
 =x_i^{N-k},\quad  \langle 0|
 { f}_{-N+k-1}  {\bar f}(y_{i})|0\rangle =y_i^{N-k},\quad k=1,2,\dots ,
 \ee
 and  (\ref{Wick-det}), it follows that
\bea\label{Delta-N-left} \langle N|f(x_1)\cdots
f(x_n)|0\rangle  &\&=\delta_{n,N}\Delta_N(x),\quad N\in \Zb,\\
\label{Delta-N-right} \langle -N|\bar{f}(y_1)\cdots
\bar{f}(y_n)|0\rangle&\&=\delta_{n,N}\Delta_N(y),\quad N\in \Zb.
\eea

Following \cite{DJKM1},\cite{JM} we consider $\hat{GL}_\infty$
element
\begin{equation}\label{g-GL}
g=e^{h},\quad h=\sum_{i,j}h_{i,j}f_i\bar{f}_j,\quad h_{i,j}\in
\mathbb{C}
\end{equation}
Via the conjugation, $(\cdot)\to g(\cdot)g^{-1}$ , each
$g\in\hat{GL}_\infty$  acts  on the spaces $\left(\oplus_{m \in
\mathbb{Z}}\mathbb{ C}f_m\right)$ and $\left(\oplus_{m\in
\mathbb{Z}}\mathbb{ C}{\bar f}_m\right)$ as  linear
transformations \cite{DJKM1},\cite{JM}.

We suppose that the following factorization condition is valid:
\begin{equation}\label{factorization}
g=g_+g_- ,\quad \langle 0|g^+ =\langle 0|, \quad g_-|0\rangle=
|0\rangle ,
\end{equation}
where $g_+,g_-\in \hat{GL}_\infty$.

Remark. Though, the property (\ref{factorization}) is valid for a
rather wide class of $(\ref{g-GL})$ (which includes all cases when
the sum in $(\ref{g-GL})$ is finite) , however, we do not know the
general theorem providing sufficient and necessary conditions to
have this property in case the sum in $(\ref{g-GL})$ is infinite.

Consider
\[
\langle 0|v_N\cdots v_1 g \bar{v}_1\cdots \bar{v}_N |0\rangle ,
\]
where each $v_i\in \left(\oplus_{m \in \mathbb{Z}}\mathbb{
C}f_m\right)$ and each $\bar{v}_i\in \left(\oplus_{m \in
\mathbb{Z}}\mathbb{ C}\bar{f}_m\right)$, $i=1,\dots,N$. Denoting
$w_i=(g_+)^{-1}v_ig_+\in  \left(\oplus_{m \in \mathbb{Z}}\mathbb{
C}f_m\right)$ and
$\bar{w}_i=(g_-)\bar{v}_i(g_-)^{-1}\in\left(\oplus_{m \in
\mathbb{Z}}\mathbb{ C}\bar{f}_m\right)$ we have
\[
\langle 0|v_N\cdots v_1 g \bar{v}_1\cdots \bar{v}_N |0\rangle=
\langle 0|w_N\cdots w_1  \bar{w}_1\cdots \bar{w}_N |0\rangle= \det
\langle 0|w_i\bar{w}_j |0\rangle |_{i,j=1,\dots,N}
\]
where the second equality is due to the Wick theorem
(\ref{Wick-det}). Thus
\begin{equation}\label{factoriz-Wick}
\langle 0|v_N\cdots v_1 g \bar{v}_1\cdots \bar{v}_N |0\rangle=\det
\langle 0|v_i g \bar{v}_j\cdots  |0\rangle |_{i,j=1,\dots,N}
\end{equation}

\subsection{Multi-component fermions }

One obtains the so-called $p$-component fermion formalism by
re-numerating  the above free fermions (\ref{fermions}) as follows
 \be \label{p-fermions}
f_n^{(\alpha)}:=f_{pn+\alpha-1}\ ,\quad {\bar
f}_n^{(\alpha)}:={\bar f}_{pn+\alpha-1}\ ,
 \ee
 \be \label{p-fermions-z}
f^{(\alpha)}(z):=\sum_{k=-\infty}^{+\infty}z^k f_{k}^{(\alpha)}\
,\quad {\bar
f}^{(\alpha)}(z):=\sum_{k=-\infty}^{+\infty}z^{-k-1}{\bar
f}_{k}^{(\alpha)}\ ,
 \ee
where $\alpha=1,\dots,p$. From (\ref{fermions}) we obviously have
\begin{equation}\label{p-fermions-antic}
[f_n^{(\alpha)},f_m^{(\beta)}]_+=[{\bar f}_n^{(\alpha)},{\bar
f}_m^{(\beta)}]_+=0,\quad [f_n^{(\alpha)},{\bar
f}_m^{(\beta)}]_+=\delta_{\alpha,\beta}\delta_{n,m}.
\end{equation}

Right and left vacuum vectors are respectively defined
\begin{equation}\label{p-vacuum-def}
|{\underbrace{0,\dots,0}_{p}}\rangle:=|0\rangle ,\quad
\langle{\underbrace{0,\dots,0}_{p}}|:=\langle 0|
\end{equation}
where $|0\rangle$ and $\langle 0|$ were introduced in (\ref{vak}).

As it follows from (\ref{vak})
\begin{eqnarray}\label{p-vak-r}
f_m^{(\alpha)} |0,\dots,0\rangle=0 \qquad (m<0),\qquad {\bar
f}_m^{(\alpha)}|0,\dots,0\rangle =0 \qquad (m \ge 0) , \\
\label{2-vak-l} \langle 0,\dots,0|f_m^{(\alpha)}=0 \qquad (m\ge
0),\qquad \langle 0,\dots,0|{\bar f}_m^{(\alpha)}=0 \qquad (m<0) .
\end{eqnarray}

We also introduce the states
 \be \label{p-vacuum}
  \langle  n^{(1)},\dots,n^{(p)}|:=\langle 0,0|C_{n^{(1)}}\cdots C_{n^{(p)}}
 \ee
where
 \bea
\label{p-vacuum'} C_{n^{(\alpha)}}&\&:={\bar f}^{(\alpha)}_0\cdots
{\bar f}^{(\alpha)}_{n^{(\alpha)}-1}
 \quad {\rm if }\ n^{(\alpha)}>0 \\
C_{n^{(\alpha)}}&\&:={ f}^{(\alpha)}_{-1}\cdots {
f}^{(\alpha)}_{n^{(\alpha)}}
\quad {\rm if}\ n^{(\alpha)}<0  \\
 C_{n^{(\alpha)}}&\&:=1 \quad {\rm if}\ n^{(\alpha)}=0
 \eea

 \be \label{p-vacuum-r}
    |n^{(1)},\dots,n^{(p)}\rangle:={\bar C}_{n^{(p)}}\cdots{\bar C}_{n^{(1)}}|0,0\rangle
 \ee
where
 \bea
\label{p-vacuum'-r} {\bar
C}_{n^{(\alpha)}}&\&:=f^{(\alpha)}_{n^{(\alpha)}-1}\cdots
f^{(\alpha)}_0 \quad {\rm if }\ n^{(\alpha)}>0 \\
{\bar C}_{n^{(\alpha)}}&\&:={\bar
f}^{(\alpha)}_{n^{(\alpha)}}\cdots {\bar f}^{(\alpha)}_{-1}
\quad {\rm if}\ n^{(\alpha)}<0  \\
 {\bar C}_{n^{(\alpha)}}&\&:=1 \quad {\rm if}\ n^{(\alpha)}=0
 \eea
Let us call (\ref{p-vacuum}) and (\ref{p-vacuum-r}) respectively
left and right charged vacuum vectors with the charge
$(n^{(1)},\dots,n^{(p)})$.

 We easily verify that
\begin{eqnarray}\label{2-vak-ch-1-r}
f_m^{(\alpha)} |*,n^{(\alpha)},*\rangle=0 \qquad
(m<n^{(\alpha)}),\qquad {\bar
f}_m^{(1)}|*,n^{(\alpha)},*\rangle =0 \qquad (m \ge n^{(\alpha)}) , \\
\label{2-vak-ch-1-l} \langle *,n^{(\alpha)},*|f_m^{(\alpha)}=0
\qquad (m\ge n^{(\alpha)}),\qquad \langle *,n^{(\alpha)},*|{\bar
f}_m^{(\alpha)}=0 \qquad (m<n^{(\alpha)}) ,
\end{eqnarray}
where $*$ serve for irrelevant components in vacuum vectors.

\begin{remark}\label{pWick}
 For calculations we use the Wick theorem in form
(\ref{Wick-det}). There are two ways to do it:

(1) The first one is to use (\ref{Wick-det}) just remembering that
$p$-component fermions are composed of usual ones, see
(\ref{p-fermions}).

(2) The second way is to use formula (\ref{Wick-det}) separately
for each component. Namely, to calculate the vacuum expectation
value of an operator $O$, first, we present it in form
 \be\label{O-comp-decomp}
O=\sum_i O^{(1)}_i \cdots O^{(p)}_i
 \ee

Then
 \be\label{Wick-comp-decomp}
\langle 0|O|0\rangle=\sum_i \langle 0|O^{(1)}_i \cdots O^{(p)}_i
|0\rangle=\sum_i \langle 0|O^{(1)}_i|0\rangle \cdots \langle
0|O^{(p)}_i |0\rangle
 \ee
where the Wick theorem in form (\ref{Wick-det}) is applied to each
of $\langle 0,\dots,0|O^{(\alpha)}_i|0,\dots,0\rangle$.
\end{remark}

\section{Fermionic representation for $Z_N$}

Consider the element of the Clifford algebra of the following form
\begin{equation}\label{g}
g=e^{A_1}g_2e^{A_2}g_3\cdots e^{A_{p-2}}g_{p-1}e^{A_{p-1}} ,
\end{equation}
where
\begin{equation}\label{A-for-2MM}
A_\alpha=\int\int f^{(\alpha)}(x){\bar
f}^{(\alpha+1)}(y)d\mu_\alpha(x,y),\quad \alpha=1,\dots,p-1,
\end{equation}
with measure $d\mu_\alpha(x,y)$, which we do not specify.

In (\ref{g})
\begin{equation}\label{g-alpha}
g_\alpha=e^{h_\alpha},\quad
h_\alpha=\sum_{i,j}h^{(\alpha)}_{i,j}f^{(\alpha)}_i\bar{f}^{(\alpha)}_j,\quad
h^{(\alpha)}_{i,j}\in \mathbb{C},
\end{equation}
 so that  we have
\begin{equation}\label{}
f^{(\beta)}_ig_\alpha=g_\alpha f^{(\beta)}_i,\quad \alpha\neq\beta
,\quad i\in\mathbb{Z}
\end{equation}
We also suppose that each $g_\alpha=e^{h^{(\alpha)}}$,
$\alpha=2,\dots,p-1$, may be factorized into
$\hat{GL}_\infty^{(\alpha)}$ elements $g_\alpha^+$ and
$g_\alpha^-$ as follows (see (\ref{factorization}))
\begin{equation}\label{factorization-alpha}
g_\alpha=g_\alpha^+g_\alpha^- ,\quad \langle
*,\stackrel{\alpha}{\hat{0}},*|g_\alpha^+ =\langle
*,\stackrel{\alpha}{\hat{0}},*|, \quad
g_\alpha^-|*,\stackrel{\alpha}{\hat{0}},*\rangle=
|*,\stackrel{\alpha}{\hat{0}},*\rangle
\end{equation}
where by $*$ we denote irrelevant components of a vacuum vector
(different from the component $\alpha$ marked by hats).

Now, let us notice that by (\ref{factoriz-Wick}) we have
\begin{equation}\label{det-rho}
\langle 0|\bar{f}^{(\alpha)}(y_1)\cdots \bar{f}^{(\alpha)}(y_N)
g_\alpha {f}^{(\alpha)}(x_N)\cdots
{f}^{(\alpha)}(x_1)|0\rangle=\det \left(\langle
0|\bar{f}^{(\alpha)}(y_i)g_\alpha
{f}^{(\alpha)}(x_j)|0\rangle\right)_{i,j=1,\dots,N}
\end{equation}

Now let us prove, that for special choice of functions
$\rho_\alpha$, $\alpha=2,\dots,p-1$, namely, for
\begin{equation}\label{rho-alpha}
\langle 0|\bar{f}^{(\alpha)}(y)g_\alpha
{f}^{(\alpha)}(x)|0\rangle=\rho_\alpha(y,x)
\end{equation}
we have
 \be \label{ev=Z_N}
(N!)^{p-1} \langle N,0,\dots,0,-N| g |0,0,\dots,0,0\rangle = Z_N
 \ee

Indeed, to get a non-vanishing expectation value in the left hand
side, we have to pick up only $N$-th term, $\frac{A_1^N}{N!}$, in
the Taylor series for $e^{A_1}$ (this is because $e^{A_1}$ is the
only factor of $g$ which contains the first component fermions,
and the matrix element $\langle N,*|A_1^n|0,*\rangle \equiv 0$
until $n= N$). Using the known formula  (\ref{Delta-N-left}), we
obtain, that the left hand side  of (\ref{ev=Z_N}) is equal to the
integral
\[
 \int d\mu_1(x^{(1)}_1,y^{(2)}_1)
 \dots \int d\mu_1(x^{(1)}_N,y^{(2)}_N)\Delta_N(x^{(1)})R_1 ,
\]
\[
R_1= \langle *,0,\dots,0,-N|\bar{f}^{(2)}(y^{(2)}_1)\cdots
\bar{f}^{(2)}(y^{(2)}_N)
 g_2\cdots|*,0,\dots,0\rangle
 \]
where we put $*$  on the first place  of the left and right vacuum
vectors to show that we forget about the first component fermions.
This is the first step.

Then, we have to pick up only $N$-th term,  $\frac{A_2^N}{N!}$,
when developing the next factor $e^{A_2}$. Otherwise, the vacuum
expectation values of the second component fermions vanishes. This
is because the second component fermions are in presence only in
$e^{A_1}$, $g_2$ and in $e^{A_2}$ factors of $g$, and the $g_2$ is
a sum of monomials, each of which contains equal number of
$f^{(2)}$ and $\bar{f}^{(2)}$ fermions, while $e^{A_1}$ contains
only $f^{(2)}$ , and $e^{A_2}$ contains only $\bar{f}^{(2)}$
fermions. Thus, second component fermions yields the expression
\[
 \bar{f}^{(2)}(y^{(2)}_1)\cdots \bar{f}^{(2)}(y^{(2)}_N)g_2
 {f}^{(2)}(x^{(2)}_1)\cdots {f}^{(2)}(x^{(2)}_N)
\]
which should be integrated with measures $\prod_{i=1}^N
d\mu(*,y^{(2)}_i) d\mu(x^{(2)}_i,*)$, and then substituted inside
$\langle N,0,\dots,0,-N|$ and $ |0,0,\dots,0,0\rangle$. Denoting
 \be\label{rho-ev-2}
\langle 0| \bar{f}^{(2)}(y^{(2)}_1)\cdots
\bar{f}^{(2)}(y^{(2)}_N)g_2
 {f}^{(2)}(x^{(2)}_1)\cdots
 {f}^{(2)}(x^{(2)}_N)|0\rangle=\varrho_2(y^{(2)},x^{(2)})
 \ee
(which, by (\ref{det-rho}), is equal to $ \det
\rho_2(y^{(2)}_i,x^{(2)}_j)$) we obtain that the l.h.s of
(\ref{ev=Z_N}) is equal to the integral
\[
 \int d\mu_1(x^{(1)}_1,y^{(2)}_1)
 \dots \int d\mu_1(x^{(1)}_N,y^{(2)}_N)\Delta_N(x^{(1)})
\]
 \[
 \int d\mu_2(x^{(2)}_1,y^{(3)}_1)
 \dots \int d\mu_2(x^{(2)}_N,y^{(3)}_N)
  \varrho_3(y^{(3)},x^{(3)})R_2 ,
\]
\[
R_2= \langle *,*,0,\dots,0,-N|\bar{f}^{(2)}(y^{(2)}_1)\cdots
\bar{f}^{(2)}(y^{(2)}_N)
 g_2\cdots|*,*,0,\dots,0\rangle
 \]
where we put $*$  on the first and second places  of the left and
right vacuum vectors to show that we forget about the first and
the second component fermions. This is the second step.

Then, it is easy to see that each exponential $e^{A_\alpha}$
should be replaced by their $N$-th Taylor term, otherwise the
l.h.s. of (\ref{ev=Z_N}) vanishes, it means we have
\[
 \langle N,0,\dots,0,-N| g |0,0,\dots,0,0\rangle=
\]
 \be \label{onlyA^N}
 \frac{1}{(N!)^{p-1}}
  \langle N,0,\dots,0,-N| A_1^Ng_2A_2^N\cdots
  g_{p-1}A_{p-1}^N |0,0,\dots,0,0\rangle
 \ee

  Continuing excluding step by
step third- forth- and so on component fermions, and, on the last
step, using the known formula (\ref{Delta-N-right}), we obtain
that (\ref{onlyA^N}) is equal to (\ref{Z_N}).

\;

At last we want to make the following remark

\br Insert additional factors to (\ref{g}) as follows
\begin{equation}\label{g-loop}
g=e^{A_1}g_2e^{A_2}g_3\cdots e^{A_{p-2}}g_{p-1}e^{A_{p-1}} \to
g_{\circlearrowleft} \; :=e^{A_1}g_2e^{A_2}g_3\cdots
e^{A_{p-2}}g_{p-1}e^{A_{p-1}}g_{p}g_1e^{A_{p}}
\end{equation}
where
  \be\label{g-A-p-loop}
g_\alpha=e^{\sum_{i,j}h^{(\alpha)}_{i,j}f^{(\alpha)}_i\bar{f}^{(\alpha)}_j}
,\quad {A_{\alpha}}=\int\int f^{(\alpha)}(x){\bar
f}^{(\alpha+1)}(y)d\mu_\alpha(x,y),\quad \alpha=1,\dots,p
  \ee
  where ${\bar
f}^{(p+1)}(y)\equiv {\bar f}^{(1)}(y)$. (Thus we add $g_1,g_{p}$
and $d\mu_p(x,y)$ to our collection of data, $g_{\alpha},\;
\alpha=2,\dots,p-1$ and $d\mu_\alpha(x,y),\; \alpha=1,\dots,p-1$).
Then
 \be \label{ev=grand-Z_N}
 \langle 0,0,\dots,0,0| g_{\circlearrowleft}
|0,0,\dots,0,0\rangle = \sum_{N=0}^\infty c_N
{Z}_N^{\circlearrowleft}
 \ee
 where $c_N$ are certain numbers and each $Z_N^{\circlearrowleft}$ is the following integral
 over $2pN$ variables
$x^{(\alpha)}=(x^{(\alpha)}_1,\dots,x^{(\alpha)}_N)$ and
$y^{(\alpha)}=(y^{(\alpha)}_1,\dots,y^{(\alpha)}_N)$, $
\alpha=1,\dots,p$:
 \be\label{loop-Z_N}
 {Z}_N^{\circlearrowleft}=
\int \prod_{\alpha=1}^{p}\varrho_\alpha(y^{(\alpha)},x^{(\alpha)})
\prod_{\alpha=1}^{p}
d\mu_{\alpha}(x^{(\alpha)},y^{(\alpha+1)}),\quad y^{(p+1)}\equiv
y^{(1)}
 \ee
(notice that variables $y^{(1)}$ and $x^{(p)}$ are not fixed by
(\ref{fixing-x-p-y-1}))

In (\ref{loop-Z_N})  $d\mu_{\alpha}(x^{(\alpha)},y^{(\alpha+1)})$
($\alpha=1,\dots,p-1$) are defined by (\ref{dmu-alpha}) and
 \be\label{dmu-alpha-loop}
d\mu_{p}(x^{(p)},y^{(1)}):=\prod_{i=1}^N
d\mu_{p}(x^{(p)}_i,y^{(1)}_i),
 \ee
and each $\varrho_\alpha(y^{(\alpha)},x^{(\alpha)})$ is defined by
(\ref{det-rho})-(\ref{rho-alpha}), where now $\alpha=1,\dots,p$.

Sums (\ref{ev=grand-Z_N}) and their relation to the grand
partition function of {\em closed} chains of coupled random
matrices and to integrable equations will be considered in a
forthcoming paper.
 \er

\section{Deformation of measure and relations to integrable
hierarchies}

The described deformation
\begin{equation}\label{measure-1-defor'}
d \mu_1(x,y)\to d\mu_1(x,y|{\bf t}^{(1)},n,\bar{\bf
t}^{(1)}):=x^{n_1} e^{V(x,{\bf t}^{(\alpha)})+V(x^{-1},\bar{\bf
t}^{(1)})) }d\mu_1(x,y),
\end{equation}
\begin{equation}\label{measure-p-1-defor'}
d \mu_{p-1}(x,y)\to d\mu_{p-1}(x,y|{\bf t}^{(p)},n,\bar{\bf
t}^{(p)}):=y^{n_{p}} e^{V(y,{\bf t}^{(p)})+V(y^{-1},\bar{\bf
t}^{(p)})) }d\mu_{p-1}(x,y)
\end{equation}
\begin{equation}\label{V-x-t'}
 V(x,{\bf t}^{(\alpha)})=\sum_{m=1}^\infty x^mt_m^{(\alpha)},\quad
  V(x^{-1},\bar{\bf t}^{(\alpha)})=
 \sum_{m=1}^\infty x^{-m}\bar{t}_m^{(\alpha)},\quad
 \alpha=1,p,
\end{equation}
Then, it is quite known fact that in this case $Z_N=\tau_N({\bf
t}^{(1)},\bar{\bf t}^{(p)})$, where $\tau_N({\bf t}^{(1)},\bar{\bf
t}^{(p)})$ is a tau function of the (one-component) TL hierarchy.
Indeed, one just re-writes (\ref{Z_N}) as $2N$-fold integral with
a modified measure $ d\mu^{mod}$ (the latter depends on the choice
of $\rho_\alpha$)  :
\[
 \int \prod_{i=1}^N d\mu^{mod}(x^{(1)}_i,y^{(p)}_i)
 x_i^{n_1}y_i^{n_p}e^{V(x^{-1}_i,\bar{\bf
t}^{(1)})+V(y^{-1}_i,\bar{\bf t}^{(p)}) }e^{V(x^{(1)}_i,{\bf
t}^{(1)})+V(y^{(p)}_i,{\bf t}^{(p)})}
 \Delta_N(x^{(1)})\Delta_N(y^{(p)})
 \]
 Moreover, as a function of
 ${\bf t}^{(1)},{\bf t}^{(p)},\bar{\bf t}^{(1)},\bar{\bf
 t}^{(p)}$, the integral
$Z_N({\bf t}^{(1)},{\bf t}^{(p)},n_1,n_p,\bar{\bf
t}^{(1)},\bar{\bf t}^{(p)})$ is a tau function of the coupled
two-component KP, or, the same, a tau function of the
two-component TL hierarchy, see \cite{paper1}.

Now, in addition, we consider the following deformations of
functions $\rho_\alpha$, $\alpha=2,\dots,p-1$,
\begin{equation}\label{rho-defor'}
\rho_\alpha(x,y)\to  \rho_\alpha(x,y|{\bf t}^{(\alpha)},{
n}^{(\alpha)},\bar{\bf t}^{(\alpha)}):=
\end{equation}
 \be\label{rho-ev-t}
\langle n^{(\alpha)}| e^{H^{(\alpha)}({\bf t}^{(\alpha)})}
\bar{f}^{(\alpha)}(y^{(\alpha)}_1)\cdots
\bar{f}^{(\alpha)}(y^{(\alpha)}_N)g_\alpha
 {f}^{(\alpha)}(x^{(\alpha)}_1)\cdots
 {f}^{(\alpha)}(x^{(\alpha)}_N)
 e^{\bar{H}^{(\alpha)}(\bar{\bf t}^{(\alpha)})}| n^{(\alpha)}\rangle
 \ee
where ${\bf t}^{(\alpha)}=({ t}^{(\alpha)}_1,{
t}^{(\alpha)}_1,\dots)$ and $\bar{\bf t}^{(\alpha)}=(\bar{
t}^{(\alpha)}_1,\bar{ t}^{(\alpha)}_2,\dots)$ are the deformation
parameters, and where the ``Hamiltonians'' $H_{k}^{(\alpha)},\;
k=\pm 1,\pm 2,\dots,$ are defined   by
 \be
\label{ham-2'} {H^{(\alpha)}({\bf
t}^{(\alpha)})}=\sum_{k=1}^\infty
H_{k}^{(\alpha)}t^{(\alpha)}_k,\quad \bar{H}^{(\alpha)}(\bar{\bf
t}^{(\alpha)})=\sum_{k=1}^\infty
H_{-k}^{(\alpha)}\bar{t}^{(\alpha)}_k ,\quad
H_{k}^{(\alpha)}=\sum_{n=-\infty}^{+\infty} f_n^{(\alpha)}{\bar
f}_{n+k}^{(\alpha)}
 \ee
(for future purpose we define them for $\alpha=1,\dots,p$ range).

Let us note that the expectation value (\ref{rho-ev-t}), by
definition \cite{DJKM1},\cite{JM},\cite{UT}, is a tau function of
one component TL and in our case may be denoted by
\[
\tau_{n^{(\alpha)}}({\bf t}^{(\alpha)}+[x^{(\alpha)}],\bar{\bf
t}^{(\alpha)}+[y^{(\alpha)}])
\]
where
\[
[x]:=\left( \frac{x}{1},\frac{x^2}{2},\frac{x^3}{3},\dots \right)
\]

Now let us prove that the combination of deformations
(\ref{measure-1-defor'})-(\ref{measure-p-1-defor'}) and
(\ref{rho-defor'})
 is equivalent to the replacement
\[
 \langle N,0,\dots,0,-N| g |0,0,\dots,0,0\rangle \to
\]
\begin{equation}\label{ev-tau}
\tau_N({\bf t},{\bf n},\bar{\bf t}):=\langle
N+n^{(1)},n^{(2)},\dots,n^{(p-1)},-N-n^{(p)}| e^{H({\bf t})}
ge^{\bar{H}(\bar{\bf t})}
|n^{(1)},n^{(2)},\dots,n^{(p-1)},-n^{(p)}\rangle ,
\end{equation}
where
\[
H({\bf t})=\sum_{\alpha=1}^p \sum_{k=1}^\infty H_k^{(\alpha)}
t_k^{(\alpha)},\quad \bar{H}(\bar{\bf t})=\sum_{\alpha=1}^p
\sum_{k=1}^\infty H_{-k}^{(\alpha)} \bar{t}_k^{(\alpha)}
\]
the ``Hamiltonians'' $H_{k}^{(\alpha)}$ were defined earlier  by
(\ref{ham-2'}).

Proof. Indeed, we have that each terms of type (\ref{rho-ev-2}),
namely, each
 \be\label{rho-ev}
\langle 0| \bar{f}^{(\alpha)}(y^{(\alpha)}_1)\cdots
\bar{f}^{(\alpha)}(y^{(\alpha)}_N)g_\alpha
 {f}^{(\alpha)}(x^{(\alpha)}_1)\cdots
 {f}^{(\alpha)}(x^{(\alpha)}_N)|0\rangle=\varrho(y^{(\alpha)},x^{(\alpha)}),\quad
 \alpha=2,\dots,p-1,
 \ee
is now replaced by
 \be\label{rho-ev-t'}
\langle n^{(\alpha)}| e^{H^{(\alpha)}({\bf t}^{(\alpha)})}
\bar{f}^{(\alpha)}(y^{(\alpha)}_1)\cdots
\bar{f}^{(\alpha)}(y^{(\alpha)}_N)g_\alpha
 {f}^{(\alpha)}(x^{(\alpha)}_1)\cdots
 {f}^{(\alpha)}(x^{(\alpha)}_N)
 e^{\bar{H}^{(\alpha)}(\bar{\bf t}^{(\alpha)})}| n^{(\alpha)}\rangle
 \ee
which is, by definition \cite{DJKM1},\cite{JM},\cite{UT}, a tau
function of one component TL. Due to (\ref{factoriz-Wick}) it is
equal to
\[
\det \rho_\alpha(x_i,y_j|{\bf t},{\bf n},\bar{\bf t})
\]
where
\[
\rho_\alpha(x_i,y_j|{\bf t},{\bf n},\bar{\bf t}):= \langle
n^{(\alpha)}| e^{H^{(\alpha)}({\bf t}^{(\alpha)})}
\bar{f}^{(\alpha)}(y^{(\alpha)}_i) g_\alpha
{f}^{(\alpha)}(x^{(\alpha)}_j)
 e^{\bar{H}^{(\alpha)}(\bar{\bf t}^{(\alpha)})}|
 n^{(\alpha)}\rangle
\]
 As for $\alpha=1,p$ we have
\begin{equation}\label{Delta-N-left-t}
\langle N,*|e^{H^{(1)}({\bf t}^{(1)})}f(x_1)\cdots f(x_N)
e^{\bar{H}^{(1)}(\bar{\bf
t}^{(1)})}|0,*\rangle=a_1\Delta_N(x)e^{\sum_{i=1}^N V(x_i,{\bf
t}^{(1)})+V(x_i^{-1},\bar{\bf t}^{(1)})},
\end{equation}
\begin{equation}\label{Delta-N-right-t}
\langle *,0| e^{\bar{H}^{(p)}(\bar{\bf
t}^{(p)})}\bar{f}(y_N)\cdots \bar{f}(y_1)
e^{\bar{H}^{(p)}(\bar{\bf
t}^{(p)})}|*,-N\rangle=a_p\Delta_N(y)e^{\sum_{i=1}^N V(y_i,{\bf
t}^{(p)})+V(y_i^{-1},\bar{\bf t}^{(p)})},
\end{equation}
where $a_\alpha=e^{\sum_{k=1}^\infty
kt^{(\alpha)}_k{\bar{t}^{(\alpha)}_k}}$, which contribute to the
deformation respectively of $d\mu_1$ and of $d\mu_{p-1}$. The end
of proof.

Thus, we obtain that the deformation of functions
$\rho_\alpha,\alpha=2,\dots,p-1$, and also of $d\mu_1,d\mu_{p-1}$
reduce to the fact that $Z_N$ is equal to $\tau_N({\bf t},{\bf
n},\bar{\bf t})$. It is known \cite{DJKM1},\cite{JM},\cite{KdL}
that thus constructed $\tau_N({\bf t},{\bf n},\bar{\bf t})$ is a
tau function of the coupled $p$-component KP hierarchy, or, the
same, $p$-component TL hierarchy.

\section{Conclusion} We equate the multi-integral (\ref{Z_N}) to
the fermionic expectation value (\ref{ev=Z_N}). On the one hand we
hope that the fermionic representation allows to evaluate
different magnitudes related to the matrix models, like spectral
determinants (compare with \cite{paper3}), or perturbative series
generalizing \cite{HO1},\cite{HO2}. On the other hand it allows to
incorporate the study of these integrals and related multi-matrix
models to the study of multi-component integrable hierarchies

\section*{Acknowledgements}

The authors  would like to thank T. Shiota and J. van de Leur for
helpful discussions, and  (A.O.)  thanks A. Odzijevicz for kind
hospitality during his stay in Bialystok in June 2005, which
helped stimulate ideas leading to this work.

\end{document}